\documentclass[%
 aip,
cp,  
 amsmath,amssymb,
 reprint, 
]{revtex4-2}

\usepackage{graphicx}
\usepackage{dcolumn}
\usepackage{bm}

\usepackage[utf8]{inputenc}
\usepackage[T1]{fontenc}
\usepackage{mathptmx} 
\usepackage{SIunits}  
\usepackage{color} 
\usepackage[normalem]{ulem}

\begin{document}

\title{Ab initio Circular Dichroism with the Yambo code: applications to dipeptides}

\author{Elena Molteni} 
 \email[Corresponding author: ]{elena.molteni@mlib.ism.cnr.it}
\affiliation{
  Istituto di Struttura della Materia - CNR (ISM-CNR), Division of Ultrafast Processes in Materials (FLASHit), Area della Ricerca di Roma 1, Via Salaria km 29.300, CP 10, Monterotondo Scalo, Roma, Italy 
}
\affiliation{%
Department of Physics, Università degli Studi di Milano, via Celoria 16, I-20133 Milano, Italy 
}
\affiliation{%
European Theoretical Spectroscopy Facility (ETSF, www.etsf.eu) 
}

\author{Giancarlo Cappellini}
\affiliation{%
European Theoretical Spectroscopy Facility (ETSF, www.etsf.eu) 
}
\affiliation{%
Department of Physics, Università degli Studi di Cagliari, and CNR-IOM SLACS Cagliari, Cittadella Universitaria di Monserrato, 09042 Monserrato, CA, Italy 
}

\author{Davide Sangalli}
\affiliation{
  Istituto di Struttura della Materia - CNR (ISM-CNR), Division of Ultrafast Processes in Materials (FLASHit), Area della Ricerca di Roma 1, Via Salaria km 29.300, CP 10, Monterotondo Scalo, Roma, Italy 
}
\affiliation{%
Department of Physics, Università degli Studi di Milano, via Celoria 16, I-20133 Milano, Italy 
}
\affiliation{%
European Theoretical Spectroscopy Facility (ETSF, www.etsf.eu) 
}

\date{\today} 

\begin{abstract}
Circular dichroism (CD) spectroscopy is a useful technique for characterizing chiral molecules. It is more sensitive than total absorption to molecule conformation, and it is routinely used
to identify enantiomers. 
We present here a first principles implementation of CD with application to three cyclo-dipeptides. 
Our CD approach for molecules has been integrated in the 5.0 release of the Yambo code\cite{Sangalli2019}, distributed under GPL.  
\end{abstract}

\maketitle

\section{\label{sec:intro} Introduction}

Circular dichroism (CD) spectroscopy, due to the different absorption of left {\it vs} right circularly polarized light by chiral systems, is a useful technique for characterizing chiral molecules, it can be used to identify different enantiomers, and it is also able to yield information on molecule conformation ({\it e.g.} \cite{Molteni_Symm_2021,Molteni_JPCB_2015} and refs. therein).
Chirality plays an important role in molecular recognition; therefore it is of interest in several fields, from drug discovery to catalysis to biomolecular function.
Computational investigations of the CD spectra of molecules can help in several ways: they can allow to predict which spectral regions are more sensitive to the absolute configuration (enantiomer) of a given molecule, and which ones to its conformation; if a chiral drug molecule
yields a clear CD “fingerprint” (i.e. well recognizable features in the CD spectrum), this may be used {\it e.g.} for assessing its accumulation in cells.

We report here on our implementation of CD calculations within the Density Functional Theory (DFT) framework in the Yambo code~\cite{Sangalli2019}, and on its application to computing (absorption and) CD spectra of three cyclo-dipeptides, cyclo(Glycine-Phenylalanine), cyclo(Tryptophan-Tyrosine) and cyclo(Tryptophan-Tryptophan), of which some of us had previously characterized the electronic occupied states\cite{Molteni_PCCPdipep2021}.

Cyclo-dipeptides (CDPs) or 2,5-diketopiperazines (DKP) are interesting both thanks to their biological and
pharmacological activities (such as antibacterial, antiviral, antitumoral, antioxidant)\cite{Mishra_Molecules_2017} and as possible building blocks for nanodevices\cite{Zhao_PepSci_2020,Jeziorna_CrystGrowthDes_2015} due to their multiple hydrogen bonding sites, which can potentially have a role in self-assembly\cite{Mattioli_SciRep_2020,Zhao_PepSci_2020}.  
As chiral molecules, CDPs can catalyze enantioselective reactions\cite{Ying_SciRep_2018}; they have been detected e.g. in meteorites\cite{Danger_ChemSocRev_2012} and can considered as precursors of longer peptides\cite{Barreiro-Lage_JPCL2021,Danger_ChemSocRev_2012}: they may therefore have had a role also in the "homochirality" of life, {\it i.e.} the prevalence of the L enantiomer of amino acids in proteins\cite{Danger_ChemSocRev_2012}.
Moreover, the role of molecule chirality in self-organization of cyclo-dipeptides has also attracted interest\cite{Jeziorna_CrystGrowthDes_2015}.

\section{\label{sec:meth} Methods}

Methyloxirane is first used as a reference molecule to validate the approach, and then we proceed to compute optical properties of the three cyclo-dipeptides. 
Optical absorption and circular dichroism (CD) spectra are computed within the Independent Particle (IP) approximation with the Yambo code~\cite{Sangalli2019}. 
Yambo is a plane wave code interfaced with QuantumESPRESSO (QE) \cite{QE_2017,QE_JPhysCondMat2009} that allows calculation of the optical response starting from the previously generated Kohn-Sham (KS) wave functions and energies in a plane-wave basis set. In the QE calculations the molecule (either methyloxirane or a cyclo-dipeptide) was put in a face-centered cubic (FCC) cell with lattice parameter $a=56.087$ a.u. ({\it i.e.} $\approx$ 29.68 \AA). Total energies have been calculated using norm-conserving Troullier-Martins atomic pseudopotentials\cite{TM_PsP}. The LDA approximation for the exchange-correlation potential is used for methyloxirane, while the hybrid B3LYP~\cite{B3LYP1,B3LYP2} functional is used for dipeptides. The D3 pairwise dispersion correction for van der Waals (vdW) interactions~\cite{Grimme_D3} was also included for relaxing the geometry of the three cyclo-dipeptides, and the Makov-Payne correction to the total energy was used to compute the vacuum
level, and to properly align electronic energy levels~\cite{MakovPayne}. In Yambo calculations we used  200 bands for methyloxirane, and 500 bands for the three investigated cyclo-dipeptides. In all cases the same  plane wave cutoff for wavefunctions of 45 Ha gives a good convergence of the energy levels, total energies and atomic forces in the QE runs.

In our approach, both absorption and CD are constructed starting from the matrix elements of the position operator which, for isolated molecules, we compute in real space
\begin{equation}
  \label{eq:r_dipoles}
  \mathbf{r}_{nm}=\langle\psi_{n}|\mathbf{r}|\psi_{m}\rangle
\end{equation}
where $\mathbf{r}$ is a vector of component $r_j$, with $j=x,y,z$.
We define $\omega_{nm}= (\epsilon_{n}-\epsilon_{m})/\hbar$, the excitation frequency for an electronic transition between the states $\psi_n$ and $\psi_m$.
From the latter we define the velocity dipoles
$\mathbf{v}_{nm}=\mathbf{r}_{nm}\omega_{nm}$, and the magnetic dipoles
\begin{equation}
  \label{eq:m_dipoles}
  \mathbf{m}_{mn}=\sum_{l=l_{min}}^{l_{max}} \mathbf{r}_{ml}\times\mathbf{v}_{ln}.
\end{equation}
The expression for the magnetic dipoles results from the use of the identity operator $\sum_l |l\rangle\langle l|$. 
Hence the $n$ and $m$ indexes belong to a given transition (from an occupied to an unoccupied orbital in the resonant case) while the $l$ index should run over all orbitals, i.e. $l_{min}=1$ and $l_{max}=\infty$. In practice one should verify the convergence of calculated CD spectra, in a given energy range, both on the transitions included, and on the $l$ index.

For Independent Particle (IP) absorption spectra we calculated the polarizability $\alpha$, in terms of the sum of direct transitions between Kohn-Sham eigenstates within the Fermi Golden Rule
\begin{equation}
  \label{eq_IPabs}
  \alpha_{ij}(\omega)=-4\pi \sum_{nm}
  \left(
  \frac{r^{i}_{nm}r^{j}_{mn}}{\omega-\omega_{nm}-i\gamma}
  +
  \frac{r^{j}_{nm}r^{i}_{mn}}{\omega+\omega_{nm}+i\gamma}
  \right).
\end{equation}
The IP circular dichroism signal, within linear response, is instead proportional to the  $G$-tensor\cite{Molteni_JPCB_2015,Condon1937,Barron2004}:
\begin{equation}
\label{eq:G_CD_IP}
G_{ij}(\omega) = \frac{q_e^2}{2m\hbar} \sum_{nm} \left( \frac{r^i_{nm}m^j_{mn}}{\omega_{nm} - \omega -i\gamma}  +   
\frac{m^j_{mn}r^i_{nm}}{\omega_{nm}+\omega+i\gamma}   \right)
\end{equation}

For randomly oriented chiral molecules absorption and CD are expressed as the trace of the $\alpha(\omega)$ and $G(\omega)$ tensors, respectively.

\section{\label{sec:res} Results}

\subsection{Numerical tests on R-methyloxirane and convergence in c-GlyPhe}
\label{subsect:Rmeth}

\begin{figure}[h]
\includegraphics[width=\textwidth]{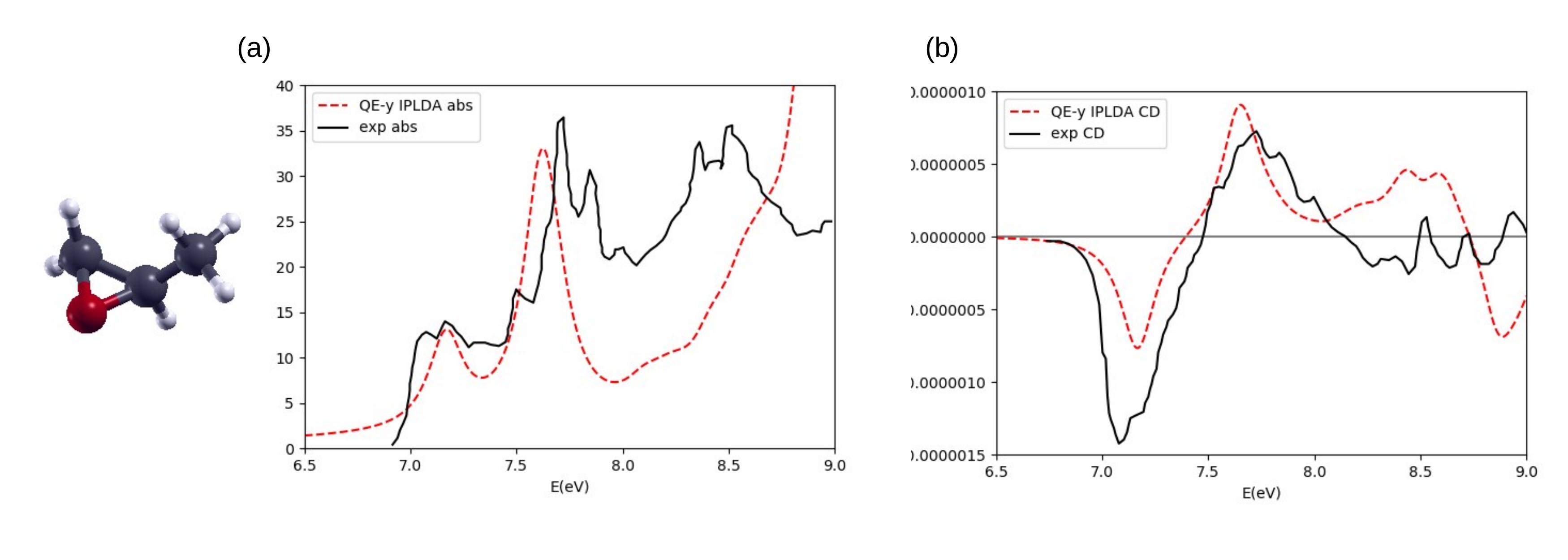}
\caption{Left panel: Geometry of R-methyloxirane. Panels (a) and (b): IP (dashed red line) absorption (panel a) and CD (panel b) calculated spectra of R-methyloxirane, obtained within DFT LDA, compared to its experimental spectra (solid black line). Calculated spectra have been shifted by +1.4 eV, and a broadening of 0.1 eV has been used.}
\label{fig:Rmeth_absCD}
\end{figure}

Methyloxirane has often been used in the literature as a benchmark molecule for CD calculations against experimental data, due to its rigidity. For flexible molecules instead, one has to take into account the fact that the different possible conformers will in general yield different CD spectra, which makes a comparison to experimental spectra not trivial. 
In Figure~\ref{fig:Rmeth_absCD} we report the geometry (left panel) of R-methyloxirane ({\it i.e.} the ``R'' enantiomer of methyloxirane) and its calculated absorption (panel a) and CD (panel b) spectra, compared to the vacuum UV experimentally measured absorption and CD spectra of the same molecule\cite{Carnell1991}. 

Our absorption and CD spectra of R-methyoxirane, calculated at IP level, reproduce well the first two features of the corresponding experimental spectra\cite{Carnell1991} provided a rigid shift of 1.4 eV is applied to calculated ones. The agreement is of the same quality of that reported in previous computational works\cite{Molteni_JPCB_2015,Varsano_PCCP_2009}. The discrepancies between theory and experiment in the high-energy part of the spectra have also been reported in the literature and they may be due to the IP approximation\cite{Molteni_JPCB_2015}.

One of the important aspects of CD implementations, is that it involves the definition of the orbital magnetic dipoles, which are ill defined in periodic boundary conditions. In our approach to the orbital magnetic dipoles, this problem appears in the presence of the terms $\mathbf{r}_{nn}$ in
eq.~\eqref{eq:m_dipoles}. In isolated systems this is not an issue, since $\mathbf{r}_{nn}$ can be directly evaluated in real space. On the other hand in extended systems the position operator is ill defined and only components between non degenerate states can be computed, via the evaluation in reciprocal space of $\mathbf{v}_{nm}$, and later using $\mathbf{r}_{nm}=\mathbf{v}_{nm}/\omega_{nm}$. Instead $\mathbf{r}_{nm}=0$ must be imposed if $\omega_{nm}<E_{thresh}$.
Here we have verified, by comparing CD spectra of R-methyloxirane obtained by computing dipoles either in real space or in reciprocal space (data not shown) that the $\mathbf{r}_{nn}$ dipoles have a negligible effect on computed CD spectra. This suggests that our approach, here implemented for molecules computing dipoles in real space, may be successfully extended to the case of solids, where the G space approach is generally used.

\begin{figure}[h]
\includegraphics[width=\textwidth]{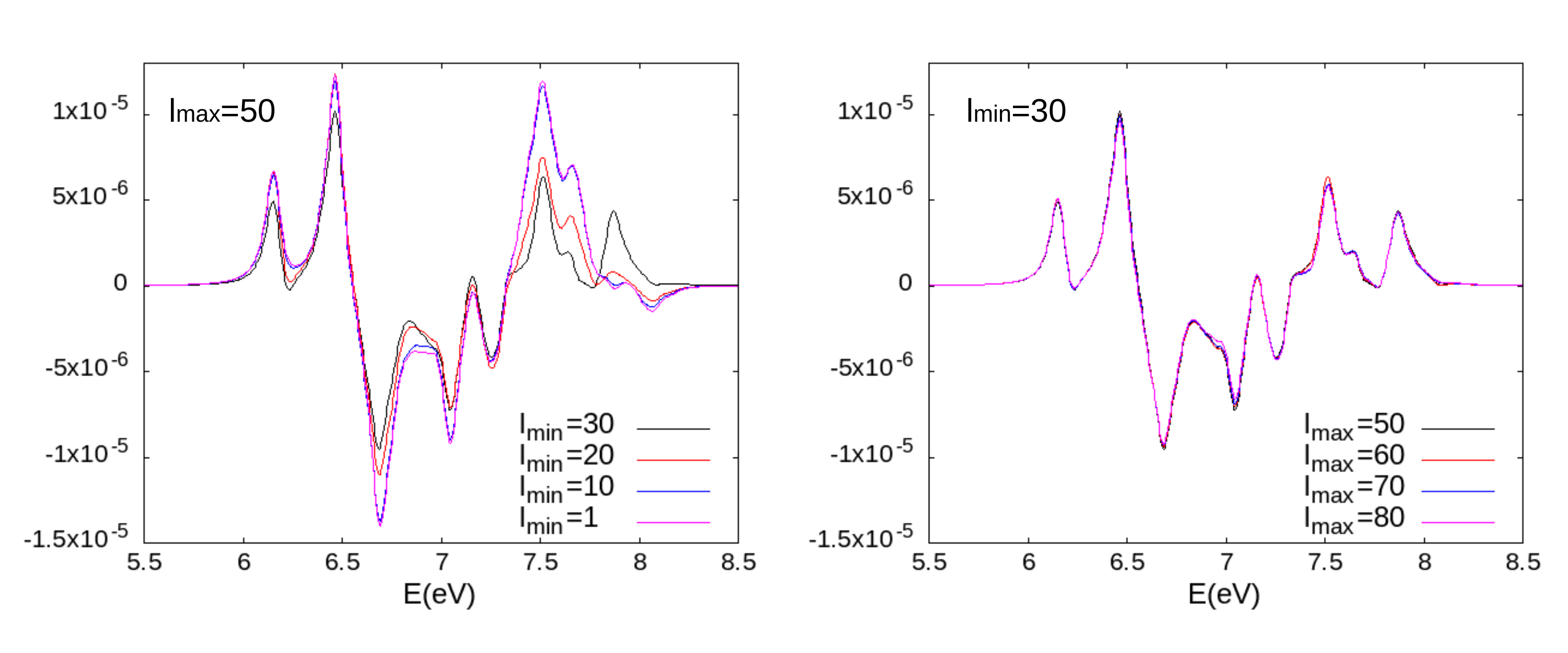}
\caption{Independent Particle CD spectra of the c-GlyPhe dipeptide, obtained with different values for the the Yambo \texttt{DipBands} keyword, used to converge the identity resolution entering the definition of the magnetic dipoles $\mathbf{m}$. Convergence is verified varying the value of $l_{min}$ (left panel), and $l_{max}$ (right panel) independently. The molecule has 39 occupied states.}
\label{fig:CD_dipbands}
\end{figure}

After validating on R-methyloxirane our computational scheme for (absorption and) circular dichroism spectra, we first consider the lowest energy conformer (see also discussion in the next section) of cyclo(Glycine-Phenylalanine) to verify the convergence of calculated CD spectra on the number of states used to resolve the identity in the definition of the magnetic dipole matrix elements (see discussion in the Methods section, $l$ index in eq.~\eqref{eq:m_dipoles}). Convergence results are presented in Fig.~\ref{fig:CD_dipbands}. The CD spectrum of c-GlyPhe is reported at fixed number of transitions, using in Eq.~\eqref{eq:G_CD_IP} 10 occupied states (index $m$ ranging from 30 to 39) and 11 empty states (index $n$ ranging from 40 to 50), while varying the number of states included in the sum over the $l$ index. In the left panel we consider the convergence with respect to the value of $l_{min}$ (index in the occupied states), while in the right panel we consider the convergence with respect to the value of $l_{max}$. The two parameters, i.e. number of transitions and number of states included in the resolution of the identity, are controlled independently in the Yambo input file {\it via} the use of the two variables \texttt{BSEbands} and  \texttt{DipBands} respectively. Magnetic dipoles, and hence CD spectra, are weakly sensitive to the range of empty states used in the expression of $\mathbf{m}_{mn}$, while the convergence on occupied states is less trivial, requiring the inclusion of occupied states down to state 10 (the molecule has 39 occupied states) for a reasonable spectrum. Here the CD spectra are computed via the real space procedure for the dipoles, since the reciprocal space dipoles would require a direct evaluation of the commutator with the non local part of the b3lyp exchange and correlation potential.

\subsection{CD spectra of Cyclo-dipeptides}
\label{subsect:3dipep}

\begin{figure}[h]
\includegraphics[width=\textwidth]{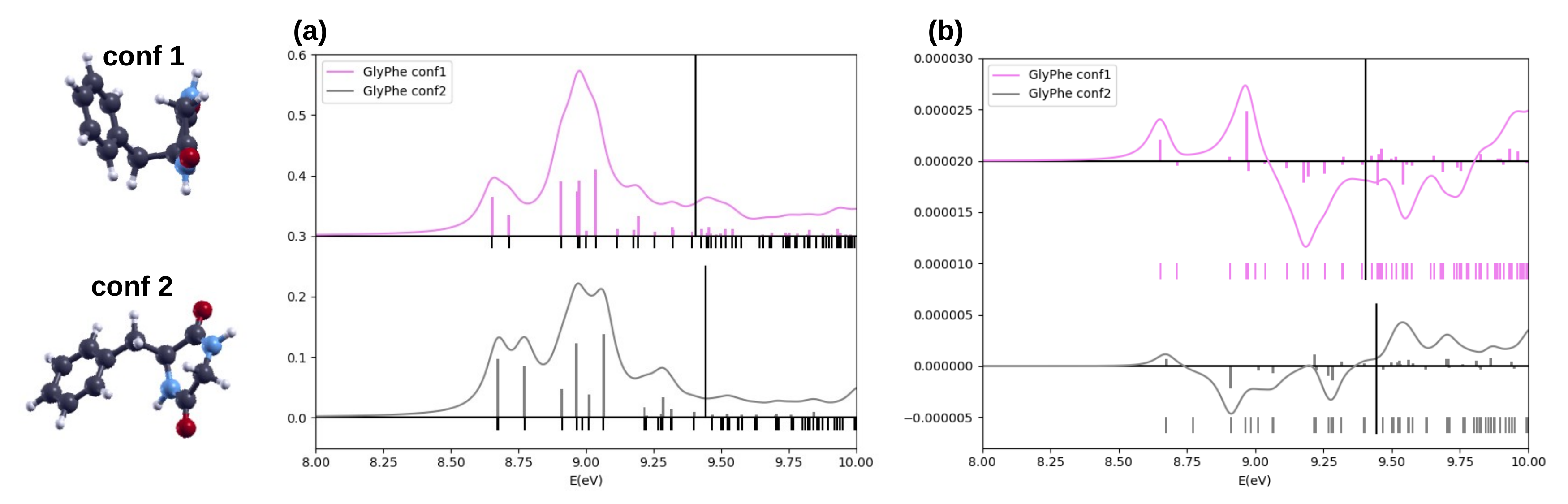}
\caption{Left panel: geometry of the chosen low-energy conformers of c-GlyPhe. Panel (a): Independent Particle (IP) B3LYP absorption spectra of c-GlyPhe conformer 1 (magenta) and conformer 2 (gray). Panel (b): IP B3LYP circular dichroism (CD) spectra of the same two conformers (same color codes). Vertical black lines indicate E(vacuum level) - E(HOMO) for the two conformations. A shift of +2.5 eV has been applied to absorption and CD spectra, and a broadening of 0.05 eV was used.}
\label{fig:GPspectra}
\end{figure}

\begin{figure}[h]
\includegraphics[width=\textwidth]{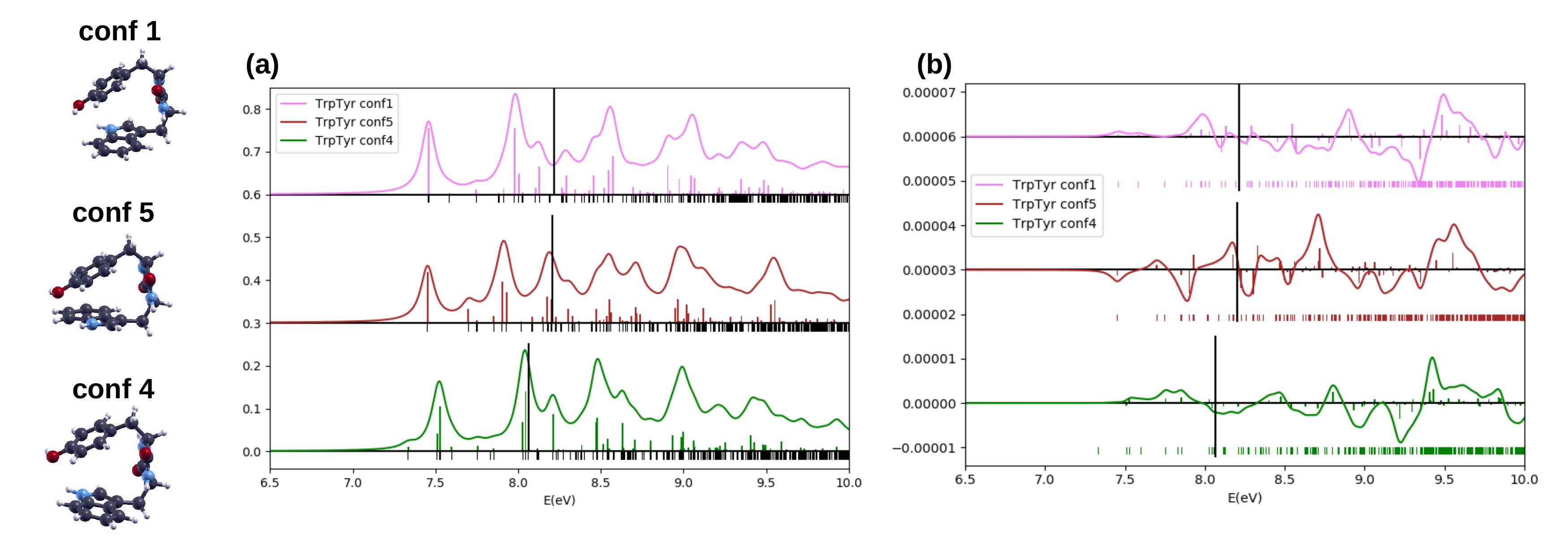}
\caption{Left panel: geometry of the chosen low-energy conformers of c-TrpTyr. Panel (a): Independent Particle (IP) B3LYP absorption spectra of c-TrpTyr conformer 1 (magenta), conformer 5 (dark red) and conformer 4 (green). Panel (b): IP circular dichroism (CD) spectra of the same three conformers (same color codes). Vertical black lines indicate E(vacuum level) - E(HOMO) for the three conformations. A shift of +2.5 eV has been applied to absorption and CD spectra, and a broadening of 0.05 eV was used.}
\label{fig:TrpTyr_spectra}
\end{figure}

\begin{figure}[h]
\includegraphics[width=\textwidth]{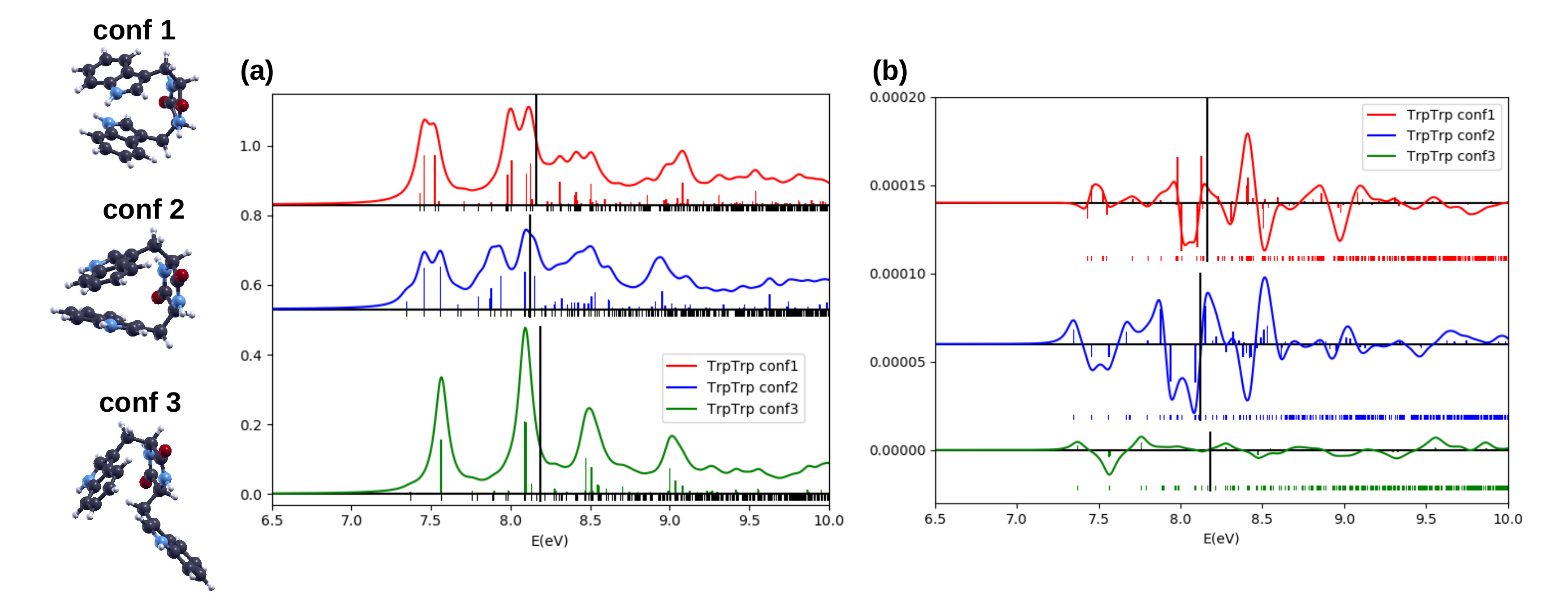}
\caption{Left panel: geometry of the chosen low-energy conformers of c-TrpTrp. Panel (a): Independent Particle (IP) B3LYP absorption spectra of c-TrpTrp conformer 1 (red), conformer 2 (blue) and conformer 3 (green). Panel (b): IP circular dichroism (CD) spectra of the same three conformers (same color codes). Vertical black lines indicate E(vacuum level) - E(HOMO) for the three conformations. A shift of +2.5 eV has been applied to absorption and CD spectra, and a broadening of 0.05 eV was used.}
\label{fig:TrpTrp_spectra}
\end{figure}

We now report results on three cyclo-dipeptides with aromatic sidechains, namely cyclo(Glycine-Phenylalanine), cyclo(Tryptophan-Tyrosine) and cyclo(Tryptophan-Tryptophan).
At a difference with the above-discussed methyloxirane, the three chosen cyclo-dipeptides (c-GlyPhe, c-TrpTyr, c-TrpTrp) display some flexibility; therefore, also in view of possible comparisons to experimentally measured absorption and/or CD spectra, one should look for the most stable conformers, which are expected to be the most abundant ones in experiments, apart from possible effects of the experimental conditions (solvent vs. gas phase, temperature, etc.). For each of the three investigated cyclo-dipeptides, therefore, we have considered the lowest energy gas phase conformers as obtained by some of us in a previous work through a tight-binding conformational search, followed by geometry optimization within B3LYP DFT\cite{Molteni_PCCPdipep2021}.
These conformers are shown in the left panels of Figs.~\ref{fig:GPspectra}, ~\ref{fig:TrpTyr_spectra} and ~\ref{fig:TrpTrp_spectra} for c-GlyPhe, c-TrpTyr and c-TrpTrp respectively, nearby the IP absorption and CD spectra.

In the IP approximation excited states correlations, which would lower the optical gap with respect to the electronic gap $E_{LUMO} - E_{HOMO}$, are neglected.  
On the other hand, the electronic gap $E_{LUMO} - E_{HOMO}$ is underestimated at the B3LYP level. The two mentioned effects are of opposite sign, therefore the IP-B3LYP-calculated energy position of the absorption onset can be either over- or underestimated with respect to the experimentally observed one, depending on which of the two effects is larger for the specific molecule under study. In the analysis of the spectra of R-methyloxirane, the sum of the two errors results in an underestimation of the optical gap. A simple rigid shift of $+1.4$~eV to the calculated IP spectrum was enough to give a very good description of CD.
For cyclo-dipeptides  instead we fix the HOMO - LUMO gap with a shift of +2.5 eV, based on a previous study~\cite{Molteni_PCCPdipep2021}, where measured photoemission spectra (PES) are carefully compared with {\it ab initio} simulations. We are thus left with an overall overestimation of the position of the optical gap. A direct comparison of IP results to experimentally measured spectra is not the main goal of the present work. This is why for all the three investigated cyclo-dipeptides we did not look for the additional shift needed to match the position of their optical spectra. In the case of the widely studied Trp amino acid\cite{Catalan_PCCP_2016,Hazra_JMCC_2014}, and also of the c-TrpTrp dipeptide\cite{Tao_NatComm_2018}, experimentally measured absorption spectra display a strong absorption peak in the 4 to 5 eV energy region, whose maximum lies at lower energy with respect to our IP calculated spectra, even before applying the above-mentioned +2.5 eV shift to them. An investigation of absorption and CD spectra of these molecules beyond IP level may be the subject of further works.

On the other hand the calculated IP spectra reported here yield information on the contribution of individual state to state transitions to the intensity of absorption and CD peaks {\it via} the dipole matrix element between the two involved states, whose intensity can be strongly affected by the localization of these states (including the possible absence of specific peaks corresponding to dipole-forbidden transitions). This information is interesting when comparing spectra of different molecular conformers, and it adds to our knowledge of the electronic properties of these cyclo-dipeptides, with respect to the simple picture obtained by DFT electronic densities of states\cite{Molteni_PCCPdipep2021} which only depend on the energy distribution of electronic states. Having aligned the transition energies with the position of the energy level measured in photo-emission, helps in mapping a peak to the associated occupied and empty states.

In Fig.~\ref{fig:GPspectra} panels (a) and (b) we compare Independent Particle (IP) B3LYP optical absorption and electronic circular dichroism (CD) spectra of conformers 1 and 2 (left panel of the same figure) of the c-GlyPhe peptide, obtained with the Yambo code from QE KS wavefunctions. 
In spite of the quite similar energy distribution of electronic levels~\cite{Molteni_PCCPdipep2021} for the two considered conformers of c-GlyPhe, absorption spectra (panel (a)) display some degree of conformational sensitivity: the main features lie in both cases around 8.7 eV and around 9 eV, but their relative intensities and detailed shapes are different for the two conformers: in conformer 2, at a difference with conformer 1, each of these two features is splitted into two peaks.
Conformational sensitivity is even more pronounced in CD spectra (panel (b)): here corresponding spectral features for the two conformers have in several cases opposite signs, thus yielding very different dichroism spectra. This strong conformational dependence of CD spectra has already been reported in the literature for single amino acids (see \cite{Molteni_JPCB_2015} and refs therein).
In c-GlyPhe conformation 1 (magenta curve in panel (a) of Fig. \ref{fig:GPspectra}) six out of the first (in energy order) seven transitions between Kohn-Sham states (short vertical black ticks below the absorption spectrum), namely all of them except the sixth one, {\it i.e.} the HOMO - (LUMO+2) one, give non-negligible contributions (vertical magenta ticks) to the Independent Particle absorption spectrum. In particular, the first and second transitions contribute to the absorption peak at $\approx$ 8.7 eV, while the 3rd, 4th, 5th and 7th transitions contribute to the absorption peak at $\approx$ 9 eV. The fact that most low-energy transitions give non-negligible contributions to the IP absorption spectrum is in agreement with the observation that most of the highest occupied and lowest unoccupied electronic states in this system, as discussed by some of us in a previous work\cite{Molteni_PCCPdipep2021}, are not localized in a specific part of the molecule (in that case low intensity contributions would be expected to occur for transitions between pairs of states localized on separated and relatively ``far'' parts of the molecule). Also in c-GlyPhe conformation 2 (grey curve in panel (a) of Fig. \ref{fig:GPspectra}) six out of the first seven transitions (short vertical black ticks) are bright (vertical grey ticks). Again only the HOMO - (LUMO+2) transition, which in this case is the fifth one in energy order, is dark. For both conformers all the mentioned bright low energy transitions have either the LUMO or the LUMO+1 as ``final'' conduction state. 
Only a subset of the optically bright transitions give non-negligible contributions to the corresponding CD spectra (panel (b) of Fig. \ref{fig:GPspectra}). In particular, in c-GlyPhe conformation 1 only the first (HOMO - LUMO) and the fourth ((HOMO-1) - (LUMO+1)) transitions give non-negligible contributions - with the same sign - to CD.
In c-GlyPhe conformation 2 instead four transitions give non-negligible contributions to CD in this energy region, namely the first (HOMO - LUMO) one with positive sign, and the third ((HOMO-1) - LUMO), sixth ((HOMO-1) - (LUMO+1)) and seventh ((HOMO-2) - (LUMO+1)) with negative sign.

In Fig.~\ref{fig:TrpTyr_spectra}, panels (a) and (b), we report the Independent Particle B3LYP absorption (a) and circular dichroism (b) spectra of the three lowest energy conformers (left panel of the same figure) of c-TrpTyr.
Although absorption spectra of the three c-TrpTyr conformers share several common features, they can be distinguished from each other. 
For c-TrpTyr conformation 1 (magenta lines in Fig. \ref{fig:TrpTyr_spectra}) the only contribution to the first IP absorption peak stems from the HOMO-LUMO transition (7.46 eV); the main contribution to the second absorption peak comes from the (HOMO-1) - LUMO transition (7.98 eV), with a less intense contribution from the (HOMO-2) - LUMO one (8 eV); the (HOMO-2) - (LUMO+1) transition (at 8.13 eV), of comparable intensity as the (HOMO-2) - LUMO one, gives rise to the shoulder of the second absorption peak.  
The mentioned transitions are the only ones giving a non-negligible contribution to IP absorption out of the 11 transitions up to 8.20 eV.
Regarding the spatial localization of the electronic states, as obtained by QE DFT B3LYP calculations in our previous work\cite{Molteni_PCCPdipep2021}, for c-TrpTyr conformation 1 most of the states near the energy gap, {\it i.e.} those involved on low energy IP transitions, are localized on a specific part of the cyclic dipeptide, not spread all over the molecule, at a difference with the case of c-GlyPhe conformation 1. In particular, the two transitions giving the most intense contributions to absorption up to 8.20 eV are the HOMO-LUMO and the (HOMO-1) - LUMO one, both involving pairs of states which are localized on the same part of the dipeptide, namely on the indole ring of tryptophan. On the other hand, transitions between pairs of states with different spatial localization yield contributions with lower intensity, such as the above mentioned (HOMO-2) - LUMO and (HOMO-2) - (LUMO+1), or even negligible intensity, such as the transitions from the HOMO state, localized on the Trp indole ring, to states ranging from (LUMO+1) to (LUMO+4), with negligible electronic density on that part of the molecule. 
As for CD spectra, their mutual differences are so pronounced that no recognizable peaks common to the three conformers are present, except possibly for the features in the range from $\approx$ 9 eV to $\approx$ 10 eV.
Also for the c-TrpTyr dipeptide the stronger conformational sensitivity of CD spectra with respect to absorption ones is thus confirmed. 
The first feature in the CD spectrum of c-TrpTyr conformation 1 (magenta curve, right panel in Fig.\ref{fig:TrpTyr_spectra}) is made of two very weak positive peaks, due to the HOMO - LUMO and to the HOMO - (LUMO+1) transitions respectively. The (HOMO-1) - LUMO transition yields the most intense contribution to the second (in energy order) CD feature, also of positive sign and more intense with respect to the previous one, lying at the same energies as the second absorption feature; however, this second CD feature has contributions also from two transitions which give negligible contributions to absorption, namely the HOMO - (LUMO+4) and the HOMO - (LUMO+5) ones.

For c-TrpTrp the differences among absorption spectra of the three B3LYP lowest energy conformers (see Fig.~\ref{fig:TrpTrp_spectra}, panel (a)) are larger than those observed for the c-TrpTyr dipeptide. The conformational variability among low energy geometries of c-TrpTrp (left panel of Fig. \ref{fig:TrpTrp_spectra}), larger than that observed in c-TrpTyr (left panel of Fig.~\ref{fig:TrpTyr_spectra}), appears to be sufficient for yielding significant differences in absorption spectra. In particular, the absorption spectrum of conformer 3 (the B3LYP lowest energy conformer, green curve in panel (a) of Fig.\ref{fig:TrpTrp_spectra}) is quite different from the spectra of the other two low energy conformers, in the relative intensity of spectral features up to $\approx$ 8.3 eV, and also, remarkably, in that both the feature at $\approx$ 7.6 eV and the one at $\approx$ 8.2 eV consist here of a single intense peak originated by two almost energy-degenerate transitions, with negligible intensity for all the other transitions in that energy range, at a difference with the other two c-TrpTrp conformers, where each of these absorption features originates from several transitions of non-negligible intensity. Interestingly, conformer 3 is rather different from the other two low-energy conformers also in its geometry, suggesting a CH-$\pi$ interaction for conformer 3, rather than a $\pi$-$\pi$ one as for conformers 1 and 2, as discussed by some of us in our previous work on these cyclo-dipeptides\cite{Molteni_PCCPdipep2021}.  
If we analyze the first absorption feature of c-TrpTrp conf3 (green curve in Fig. \ref{fig:TrpTrp_spectra} panel (a)) in terms of transitions between Kohn-Sham states (short vertical black lines below the spectrum), we find that only the second and the third transition (in order of energy) yield a non-negligible intensity in the energy range up to $\approx$ 7.8 eV: they are the HOMO - (LUMO+1) and the (HOMO-1) - LUMO transitions (the two intense green vertical lines at $\approx$ 7.564 eV and at $\approx$ 7.567 eV, respectively, appearing as a single line in the Figure). 
Interestingly, the computed B3LYP wavefunctions\cite{Molteni_PCCPdipep2021} of the HOMO and LUMO+1 states are both localized on the same part of the TrpTrp dipeptide, {\it i.e.} on the indole ring of one (the same in both cases) of the two Trp amino acids. The wavefunctions of the HOMO-1 and LUMO states, on the other hand, are both localized on the indole ring of the other Trp. 
The other transitions in the same energy range, such as the first one in order of energy, HOMO - LUMO, the fourth one,  (HOMO-1) - (LUMO+1), the fifth one, HOMO - (LUMO+2), yield negligible intensities, and they correspond to pairs of electronic states localized in different regions of the molecule.
The (HOMO-2) - (LUMO+1) and (HOMO-3) - LUMO transitions, both at $\approx$ 8.09 eV, build the second intense absorption peak for c-TrpTrp conformer 3: they are, once again, transitions between pairs of electronic states localized on the same part of the molecule, {\it i.e.} the indole ring of either of the two Trp amino acids.
As for circular dichroism (Fig.~\ref{fig:TrpTrp_spectra} panel (b)), the spectra of conformers 1 and 2 of c-TrpTrp display a similar feature around 8 eV; the spectrum of conformer 3 (the lowest energy geometry) is overall less intense than the spectra of the other 2 investigated conformers. 
This latter CD spectrum (green curve in panel (b)) displays a first weak positive peak at 7.37 eV, due to the optically dark HOMO - LUMO transition. The following CD peak is more intense, of negative sign, and due to the HOMO - (LUMO+1) and (HOMO-1) - LUMO transitions, {\it i.e.} the ones involved in the first absorption peak. Then, another positive CD peak is due to the (HOMO-1) - (LUMO+1) transition, which is dark in the absorption spectrum. Finally, the (HOMO-2) - (LUMO+1) and (HOMO-3) - LUMO transitions yield a negligible contribution to the CD spectrum, at a difference with the absorption spectrum, where they yield the second (in energy order) intense peak observed.

\section{Conclusion}
In this work we have reported on our implementation of circular dichroism calculations at Independent Particle level in the Yambo code and on its application to three cyclo-dipeptides, cyclo(Glycine-Phenylalanine), cyclo(Tryptophan-Tyrosine) and cyclo(Tryptophan-Tryptophan), with some considerations on the more pronounced conformational sensitivity of CD with respect to absorption, and on the interpretation of Independent Particle spectral features in terms of both the energy position of occupied and empty frontier orbitals, and of their spatial localization. The implementation of CD calculations beyond IP level will be the subject of future works. 

In particular, an analysis of Independent Particle absorption spectra of the three investigated cyclo-dipeptides, together with the spatial localization of the electronic states involved in the optical transitions (see Results section) illustrates how similar densities of electronic energy levels can potentially yield rather different absorption spectra, due to the generally higher (lower, respectively) intensities of contributions from transitions between pairs of orbitals with a large (resp. small) spatial overlap. Moreover, the fact that the shape of circular dichroism spectra is affected by the positive or negative sign of the CD contributions corresponding to individual optical transitions, in addition to their intensity, and that not all absorption features will in general have a non-negligible CD counterpart, increases the conformational variability of CD spectra with respect to absorption ones. This renders CD spectroscopy potentially useful for conformational analysis, while at the same time requiring careful consideration when using it for identifying enantiomers of flexibile molecules.

\begin{acknowledgments}
The present work was performed in the framework the PRIN
20173B72NB research project “Predicting and controlling the fate
of biomolecules driven by extreme-ultraviolet radiation”, which
involves a combined experimental and theoretical study of electron dynamics in
biomolecules with attosecond resolution. 
We moreover acknowledge financial support from IDEA-AISBL, Bruxelles, Belgium and from European Union project MaX Materials design at the eXascale H2020-EINFRA-2015-1 (Grant Agreement No. 824143).
\end{acknowledgments}

\nocite{*}
\bibliography{procNano2021}

\end{document}